\documentclass{article}
\usepackage{emulateapj}
\usepackage{apjfonts}
\usepackage{epsf}

\slugcomment{Submitted to ApJ Letters November 5, 1999 }

\begin{document}

\title{Formation of Short-Period Binary Pulsars in Globular Clusters}

\author{Frederic A.~Rasio\altaffilmark{1}, Eric D.~Pfahl, and
        Saul Rappaport}
\affil{Department of Physics, MIT, Cambridge, MA 02139}
\altaffiltext{1}{Sloan Research Fellow; rasio@mit.edu.}

\begin{abstract}
We present a new dynamical scenario for the formation
of short-period binary millisecond pulsars
in globular clusters. Our work is motivated by the recent observations 
of 20 radio pulsars in 47~Tuc. In a dense cluster
such as 47~Tuc, most neutron stars acquire binary companions
through exchange interactions with primordial binaries.
The resulting systems have semimajor axes in the range $\sim 0.1-1\,$AU
and neutron star companion masses $\sim 1-3\,M_\odot$.
For many of these systems we find that, when the companion
evolves off the main sequence and fills its Roche lobe, the subsequent
mass transfer is dynamically unstable. This leads
to a common envelope phase and the formation of short-period
neutron star -- white dwarf binaries. For a significant fraction
of these binaries,
the decay of the orbit due to gravitational radiation will be followed
by a period of stable mass transfer
driven by a combination of gravitational radiation and tidal heating
of the companion. The properties of the resulting short-period
binaries match well those of observed binary pulsars
in 47~Tuc.
\end{abstract}

\keywords{celestial mechanics, stellar dynamics --- clusters: globular ---
pulsars: general --- stars: neutron}

\section{Introduction}

Twenty millisecond radio pulsars have now been observed
in the globular cluster 47~Tuc (Camilo et al.\ 2000; Freire et al.\ 2000).
This is by far the largest sample of radio pulsars known in any globular
cluster. Accurate timing solutions, including positions in the cluster,
are known for 14 of the pulsars.
These recent observations provide a unique opportunity
to re-examine theoretically the formation and evolution of recycled pulsars
in globular clusters.

The binary properties of the 47~Tuc pulsars are rather surprising. While
7 pulsars are single, the majority are in short-period binaries. Most of
the binaries (8 out of 13) have properties similar to those of the rare
``eclipsing
binary pulsars'' seen in the Galactic disk population (see Nice 2000 for a
review).
These systems have extremely short orbital periods, $P_b\sim1-10\,$hr,
circular orbits, and
very low-mass companions, with $m_2\sin i\sim 0.01-0.1\,M_\odot$.
The remaining 5 binaries have properties more similar to those of the bulk
disk population, with nearly-circular orbits, periods $P_b\sim1-3\,$d (near
the short-period end of the distribution for binary millisecond pulsars
in the disk) and companions of mass $m_2\sin i\simeq 0.2\,M_\odot$.

The large inferred total population of recycled
pulsars in 47~Tuc ($\sim10^3$, see Camilo et al.\ 2000) and
the high
central density of the cluster ($\rho_c\sim10^5-10^6\,M_\odot\,{\rm pc}^{-3}$,
see De Marchi et al.\ 1996; Camilo et al.\ 2000) suggest that
dynamical interactions must play a dominant role in the formation of
these systems. However, the two dynamical formation scenarios traditionally
invoked for the production of recycled pulsars in globular clusters
clearly fail to explain the observed binary properties of the
47~Tuc pulsars.

Scenarios based on {\em tidal capture\/} 
of low-mass main-sequence stars (MS) by
neutron stars (NS), followed by accretion
and recycling of the NS
during a stable mass-transfer phase, run into many difficulties.
Serious problems have been pointed out about the tidal capture process
itself (which, because of strong nonlinearities in the regime relevant to
globular clusters, is far more likely to result in a merger than in the
formation
of a detached binary; see, e.g., Kumar \& Goodman 1996; McMillan et al.\ 1990;
Rasio \& Shapiro 1991; Ray et al.\ 1987). Moreover, the basic predictions of 
tidal capture
scenarios are at odds with many observations of binaries and pulsars in
clusters (Bailyn 1995; Johnston et al.\ 1992; Shara et al.\ 1996).
It is likely that
``tidal-capture binaries'' are either never formed, or contribute
negligibly to the production of recycled pulsars.
Verbunt (1987) proposed that collisions between NS and red giants
might produce directly neutron star -- white dwarf (NS-WD) 
binaries with ultra-short periods, but
detailed hydrodynamic simulations later showed that this does not
occur (Rasio \& Shapiro 1991).

The viability of tidal capture and two-body collision
scenarios has become less relevant with
the realization over the last 10 years that globular clusters contain
dynamically significant populations of {\em primordial binaries\/} (Hut et
al.\ 1992).
Neutron stars can then acquire binary companions through {\em exchange
interactions\/}
with these primordial binaries. Because of its large cross section, this
process dominates over any kind of two-body interaction even for low
primordial binary fractions (Heggie et al.\ 1996; Leonard 1989;
Sigurdsson \& Phinney 1993).
In contrast to tidal capture, exchange interactions with
hard primordial binaries (with semimajor axes $a\sim0.1-1\,$AU)
can form naturally the wide
binary millisecond pulsars seen in some low-density globular clusters
(such as PSR B1310$+$18, with $P_b=256\,$d, in M53, which has the lowest
central density, $\rho_c\sim10^3\,M_\odot\,{\rm pc}^{-3}$, of any globular
cluster with observed radio pulsars; see, e.g., Phinney 1996).
When the newly acquired MS companion, of mass
$\la1\,M_\odot$, evolves up the giant branch, the orbit circularizes
and a period of {\em stable\/} mass transfer
begins, during which the NS is recycled (see, e.g., Rappaport et al.\ 1995).
The resulting NS-WD binaries
have orbital periods in the range $P_b\sim1-10^3\,$d.
However, this scenario
cannot explain the formation of recycled pulsars in binaries with
periods shorter than $\sim1\,$d. To obtain such short periods,
the initial primordial binary must be extremely hard, with $a \la 0.01\,$AU,
but then the recoil velocity of the system following the exchange interaction
would almost certainly exceed the escape speed from the shallow
cluster potential ($v_e\simeq 60\,{\rm km}\,{\rm s}^{-1}$ for 47~Tuc).

One can get around this problem by considering more carefully
the stability of mass transfer in NS-MS binaries formed through exchange
interactions.
While all MS stars in the cluster {\em today\/} have masses $\la1\,M_\odot$,
the rate of exchange interactions may very well have peaked at a time
when significantly more massive MS stars were still present.
Indeed, the NS and the most massive primordial binaries will
undergo mass segregation and concentrate in the cluster core on a time scale
comparable to the initial half-mass relaxation time $t_{rh}$. For a dense
cluster like 47~Tuc, we expect $t_{rh}\simeq 10^9\,$yr,
which is comparable to the MS lifetime of a $\simeq 2-3\,M_\odot$ star.
If the majority of NS acquired MS companions in the range of $\sim1-3\,M_\odot$
(as we find), a
drastically different evolution may follow. Indeed, in this case,
when the MS star evolves and fills its Roche lobe, the
mass transfer for many systems (depending on the mass ratio and evolutionary
state of the donor star) is {\em dynamically unstable\/} and leads to a
common-envelope
(CE) phase. The emerging binary will have a low-mass WD in
a short-period, circular orbit around the NS.
This simple idea is at the basis of the evolutionary scenario we explore
quantitatively in \S2.
A similar scenario, but starting from tidal capture binaries and applied to
X-ray sources in globular clusters, was discussed by Bailyn \& Grindlay (1987).
The possibility of forming
intermediate-mass binaries through exchange interactions was mentioned
by Davies \& Hansen (1998), who pointed out that NS retention in 
globular clusters may
also require that the NS be born in massive binaries. Among
eclipsing pulsars in the disk, at least one system (PSR J2050$-$0827)
is likely to have had an intermediate-mass binary progenitor,
given its very low transverse velocity (Stappers et al.\ 1998).

\section{Formation and Evolution of Short-Period Binaries}

We have carried out Monte-Carlo simulations
to test quantitatively a formation scenario based
on the ideas outlined in \S1. The general framework follows that used in previous
Monte-Carlo studies of binary evolution and cluster dynamics
(Di Stefano \& Rappaport 1994; Hut, McMillan, \& Romani 1992;
Joshi, Rasio, \& Portegies Zwart 2000;
Rappaport, Di Stefano, \& Smith 1994, hereafter RDS).
More details on our Monte-Carlo method and dynamical simulations,
as well as a systematic study of how
the results depend on model parameters, will be presented in 
forthcoming papers
(Joshi et al.\ 2000; Pfahl et al.\ 2000). Here we simply outline
the major steps in our simulations and we present some 
representative results: 
\smallskip

\noindent
(1) We begin with a population of primordial MS binaries and single NS,
distributed as a constant fraction of the mass density in the cluster.
Primary masses $m_1$ are selected using the Eggleton (2000) Monte-Carlo
representation of the Miller \& Scalo (1979) IMF (see eq.~[1] of RDS).
The secondary mass
is chosen so that the probability distribution for the
binary mass ratio $q=m_2/m_1$ is $p(q)\propto q^{1/4}$ (Abt \&
Levy 1978).
Initial binary orbital periods are distributed uniformly in $\log P$
over the interval $\sim10^{-1}-10^8\,$d. Eccentricities are
generated from a thermal distribution, $p(e_i)=2e_i$.
All neutron stars have
mass $m_{NS}=1.4\,M_\odot$. For definiteness we start with $5\times10^6$
binaries
and $10^4$ NS in a cluster containing a total of $10^7$ stars.
These numbers affect only the overall normalization of our results.
\smallskip

\noindent
(2) Binaries and NS undergo mass segregation and enter
the cluster core in a time $t_s$, distributed according to
$p(t_s) = (1/t_{sc}) \exp ( - t_s / t_{sc})$,  where the characteristic
 time $t_{sc} \simeq 10 (m_f/m_t) t_{rh}$ for objects of mass $m_t$
drifting through field stars of average mass $m_f$.
This simple analytic law fits very well the results of detailed dynamical
simulations of mass segregation (Fregeau, Joshi, \& Rasio 2000). We fix 
$t_{rh}=10^9\,$yr in this paper.
Binaries whose primaries evolve off the MS before entering the 
cluster core are removed from the simulation.
\smallskip

\noindent
(3) We assume a fixed
core density $n_c=10^5\,{\rm pc}^{-3}$, core radius $r_c=0.5\,$pc,
and 3D velocity dispersion $v_c=15\,{\rm km}\,{\rm s}^{-1}$
(meant to be average values over the evolution of the cluster).   
The fraction of the core density in single NS increases slowly
as NS drift into the core, and reaches a maximum of 
about 5\% of the core density at $\sim4\times10^9\,$yr before 
exchange interactions begin to deplete their population significantly. 
From the numbers of binaries and NS in the core,
we compute the time for each binary to have a strong interaction. 
Here a strong interaction is defined to have a distance
of closest approach $<a_i (1+e_i)$, where $a_i$ and $e_i$ are
the initial binary semimajor axis and eccentricity.
Soft binaries, with binding energies $<m_fv_c^2/2$, are assumed to be
disrupted by the interaction. Very hard binaries will have recoil
velocities $v_{rec} > v_e=60\,{\rm km}\,{\rm s}^{-1}$ and will be
ejected from the cluster. Here we approximate the
results of scattering experiments by taking $v_{rec} \simeq 0.1 v_b$, where
$v_b$ is the binary orbital velocity.
Disrupted and ejected binaries, as
well as those whose primaries evolve off the MS before interacting,
are removed from the simulation.
\smallskip

\noindent
(4) Of the binaries that survive their first strong
interaction, $1/2$ are assumed to form a new NS-MS binary through
exchange (most of the rest will experience a direct stellar collision
and merger and we do not follow their evolution). For simplicity
we assume that
the less massive member of the original binary is always ejected in
the exchange interaction (cf.\ Heggie et al.\ 1996).
We approximate  the results of scattering experiments by taking
the final semimajor axis $a_f \simeq a_i$ and by generating a 
final eccentricity from a thermal distribution. 
\smallskip

\noindent
(5) We now calculate the evolution of the newly formed NS-MS binaries.
When the primary evolves off the MS, the orbit is assumed to circularize
(conserving total angular momentum). We then test for the stability
of mass transfer  when the primary fills its Roche lobe
(using eq.~[33] of Rappaport et al.\ 1983, with $\xi_{ad}$ adapted from
new, unpublished results of P.~Podsiadlowski; see also
Kalogera \& Webbink 1996). We find that,
with the parameters adopted above, about 50\% of the systems enter a
CE phase. The outcome of the CE phase is calculated using the standard
treatment, with the efficiency parameter $\alpha_{CE}=0.5$ (defined
as in eq.~[2] of RDS). The WD (core)
mass is calculated from the progenitor mass and Roche lobe radius
as in RDS.
\smallskip

\noindent
(6) A significant fraction of these NS-WD binaries will undergo further
evolution driven by gravitational radiation. For orbital periods
$\la 8\,$hr, the companion will be filling its Roche lobe in less
than $\sim10^{10}\,$yr and a second phase of mass transfer will occur.
For WD masses $\la 0.4\,M_\odot$ the mass transfer is stable and
the evolution can be calculated semi-analytically
using standard methods and assumptions (e.g., Li et al.\ 1980;
Rappaport et al.\ 1987). We track the accretion rate and spin-up
of the NS during the mass-transfer phase and we terminate the evolution
when the NS spin period reaches a randomly chosen value in the range 
$2-5\,$ms (at which point the radio
pulsar emission is assumed to turn on and stop the accretion flow).
Results for a 
typical system are illustrated in Fig.~1, and for the
entire population in Fig.~2.
In its simplest version, the calculation assumes that the NS companion
remains degenerate during the entire evolution. In an effort to better
match the observed properties of the 47~Tuc binaries, we have also
considered a modified 

\medskip
\epsfxsize=8truecm
\epsfbox{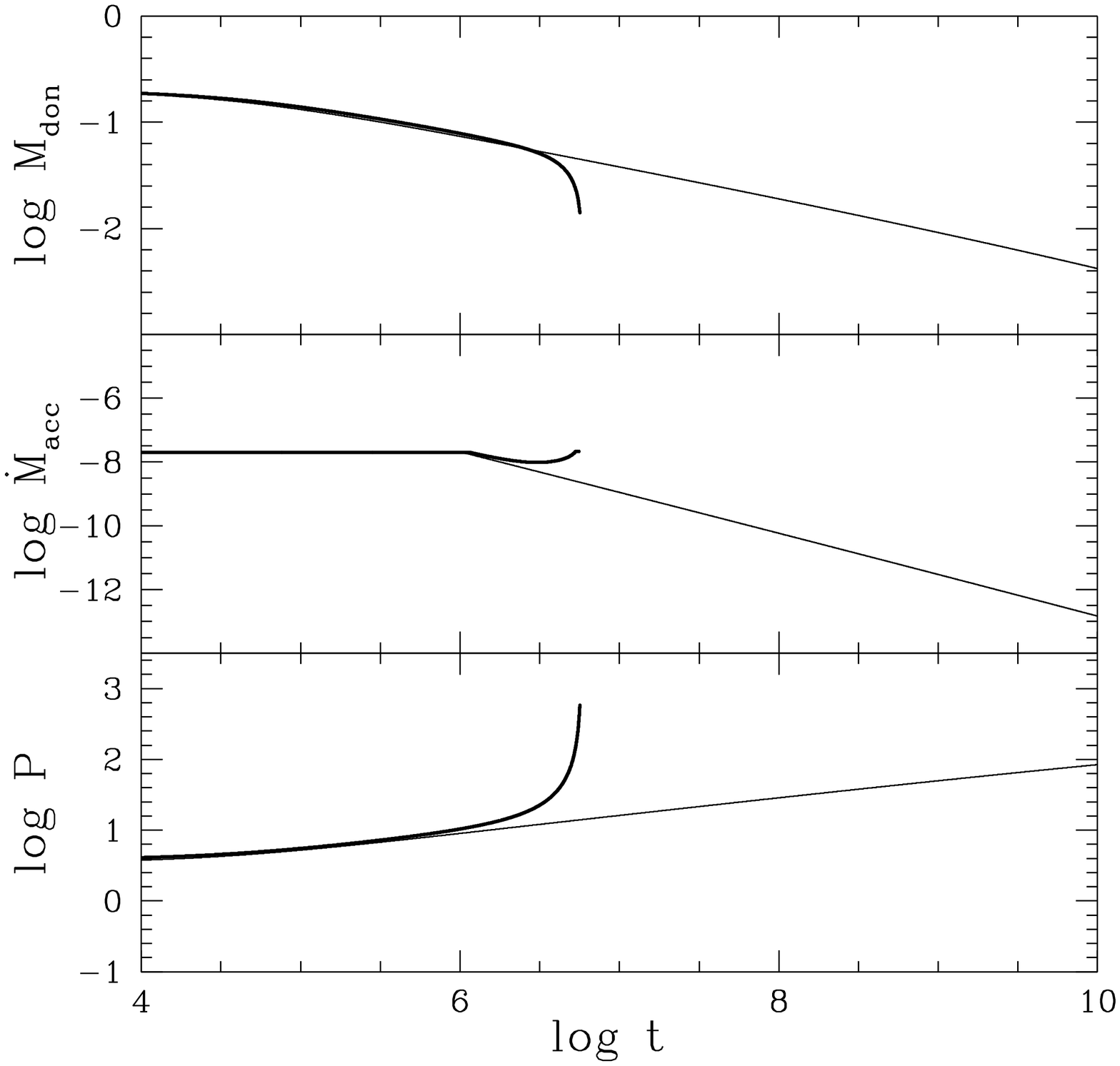}
\figcaption[fig1.ps]{Evolution of one representative NS-WD binary
driven by gravitational radiation only (thin solid lines) and by a combination
of gravitational radiation and tidal heating (thick solid lines).
Here time $t$ is in yr, the orbital period $P$ is in minutes, the mass
accretion rate $\dot M_{acc}$ (onto the NS) is in $M_\odot\,{\rm yr}^{-1}$,
and the companion (donor) mass $M_{don}$ is in $M_\odot$.
  \label{fig1}}
\bigskip
\epsfxsize=8truecm
\epsfbox{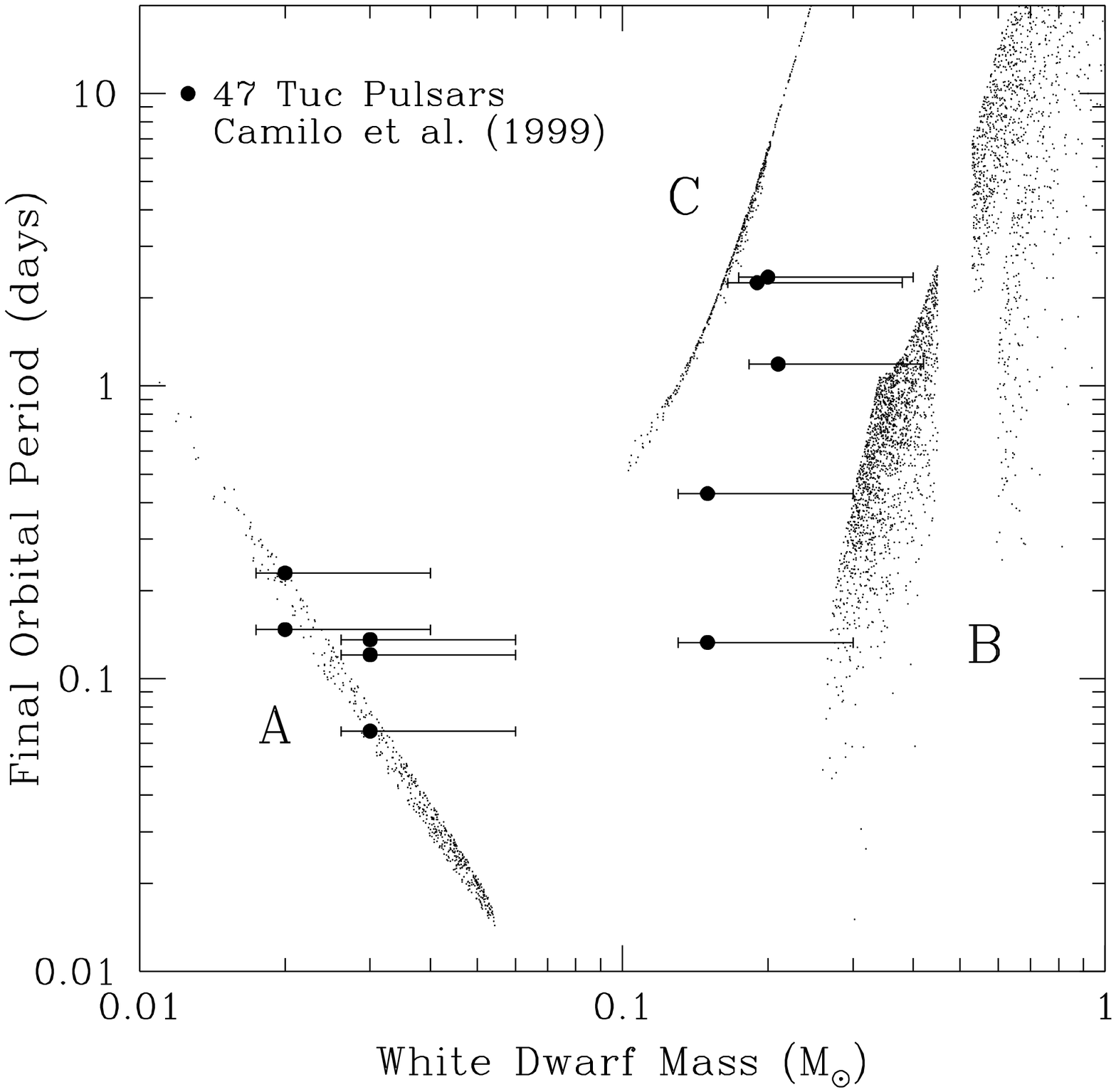}
\figcaption[fig2.ps]{
Results of our initial population synthesis study for
binary millisecond pulsars in 47~Tuc.  Each small dot represents a binary
system in our simulation, while the filled circles are the 10 binary 
pulsars in 47~Tuc with
well measured orbits (the error bars extend from the minimum companion
mass to the 90\% probability level for random inclinations). 
There are 3 principal groups of simulated binaries.  
Systems in the
diagonal band on the left (A) are binaries that decayed via
gravitational radiation to very short orbital periods ($\sim\,$mins), then
evolved with mass transfer back up to longer periods under the influence of
both gravitational radiation and tidal heating.  
The large group labeled B on the right
contains NS-WD binaries which had insufficient time to decay to Roche-lobe
contact via the emission of gravitational radiation.  
The NS in this group are not likely to be
recycled since they may not have accreted much
mass during the CE phase.
Finally, the systems lying
in the thin diagonal band toward longer periods (C) are those in which the mass
transfer from the giant or subgiant to the NS would be stable.  These have not
been evolved through the mass transfer phase; the mass plotted is simply
that of the He core of the donor star when mass transfer commences.  There
are many more systems in this category that have longer periods but lie off
the graph.  The numbers in the three groups are: $N_A\sim1000$, $N_B\sim2400$, 
and $N_C\sim3500$.
  \label{fig3}}
\bigskip

\noindent
evolution in which the companion becomes  tidally
heated and non-degenerate (but still modeled as a simple $n=3/2$
polytrope), as appears to be the case in many eclipsing
binary pulsars (Applegate \& Shaham 1994; Nice 2000). We adopt a synchronization
time $t_{syn}=6\times10^4\,$yr and a (magnetically driven) asynchronism
$\Delta=|\Omega_s-\Omega_b|/\Omega_b=0.3$, in agreement with the
values suggested by Applegate \& Shaham (1994) for PSR B1957$+$20.
Note, however, that in our scenario the companion is initially
degenerate, while Applegate \& Shaham (1994) start with a low-mass
MS companion.

\section{Discussion}

Our scenario provides a natural way of explaining the large
number and observed properties of short-period binary pulsars
in 47~Tuc.
Although quantitatively the predicted properties of the final binary
population depend on our parametrization of several
uncertain processes (such as CE evolution and tidal heating),
the overall qualitative picture is remarkably robust. We find that,
quite independent of the details of our various assumptions and
choices of parameters, exchange interactions {\em inevitably\/} form
a large population of NS-MS binaries that will go through a CE phase.
The only way for a globular cluster to avoid forming such a population
would be to start with a very low primordial binary fraction, a
very small number of retained NS, or to have a very long relaxation
time $t_{rh}\ga10^{10}\,$yr, such that all MS stars with masses
$\ga1\,M_\odot$ evolve before the rate of exchange interactions becomes
significant. A large fraction of the post-CE NS-WD binaries cannot
avoid further evolution driven by gravitational wave emission, with
the companion ultimately driven to a very low mass
$m_2\sim 10^{-2}\,M_\odot$.

A limitation of this preliminary study is that we do not take
into account multiple interactions. The average collision time for a
hard binary
with component masses $m_1=1.4\,M_\odot$ and $m_2=0.1\,M_\odot$
and orbital period $P_d\,$d, in a cluster of density
$n_5\,10^5\,{\rm pc}^{-3}$ and 1D velocity dispersion $\sigma_{10}\,10\,
{\rm km}\,{\rm s}^{-1}$, is $t_{coll}\simeq 10^{10}\,{\rm yr}\,n_5^{-1}\,
\sigma_{10}\,P_d^{-2/3}$. Thus, in Fig.~2, all binaries with periods
$\ga1\,$d will be affected by further interactions if they reside in the
cluster core (consistent with the positions of the wider binary
pulsars well outside the core of 47~Tuc; see Rasio 2000). 
For a small fraction of systems that undergo multiple
interactions, the NS may acquire a new MS companion that will be evolving
before the next interaction, thereby leading to essentially the same 
type of evolution already considered in \S2. 
Another small fraction may liberate the NS.
This could be an important channel for forming single millisecond pulsars
in globular clusters, although complete evaporation of a low-mass companion
 (as proposed for the disk population of single millisecond pulsars
by Kluzniak et al.\ 1988) is another possibility. However,
in most cases, multiple interactions will
lead to the direct collision of the NS with a MS star, especially if
the cluster core is dominated by binaries and resonant binary-binary encounters
are frequent (Bacon et al.\ 1996). The outcome of such collisions is highly
uncertain (see Fryer et al.\ 1996 for a recent discussion).
Note that, if, as we suggest, recycled pulsars in short-period binaries
have progenitors that went through a CE phase, then the NS must be
able to survive inside the envelope of a low-mass giant without
hypercritical accretion and subsequent collapse to a black hole.
This is in agreement with the results of Fryer et al.\ (1996).

In addition to explaining the
short-period binary millisecond pulsars in 47~Tuc,
our scenario for the evolution of NS-WD binaries driven by gravitational
radiation and tidal heating may be relevant to
eclipsing binary pulsars in the disk population,
as well as to short-period X-ray binaries such as 4U 1820$-$30, 4U1850$-$087,
4U1626$-$67, 4U1916$-$053, and SAX J1808$-$3658. 
In particular, 4U 1820$-$30 in the
globular cluster NGC 6624, with an orbital period of $\sim11\,$min, may be
the prototypical NS-WD system observed during the short-lived, bright
X-ray phase of its evolution (Rappaport et al.\ 1987). The stable,
super-Eddington
mass-transfer phase for these systems lasts typically $\sim10^6-10^7\,$yr
(see\ Fig.~1). Since these
recycled pulsars do not have long-lived X-ray binary progenitors,
our scenario naturally avoids a ``birthrate problem''
(Kulkarni, Narayan, \& Romani 1990).

While the properties of 5 ``eclipsing binary pulsars'' 
are clearly well explained by our
scenario (group A in Fig.~2), the other group of 5 binaries
in 47~Tuc with companion 
masses $\sim 0.2\,M_\odot$ lie distinctly toward
smaller masses than the simulated systems in the left part of group~B 
(with He WDs).
We speculate that these pulsars may in fact have
evolved from the group of systems with stable mass transfer from a 
$\sim1\,M_\odot$ subgiant 
to a NS, which have orbital periods at the start of mass transfer in
the range $\simeq 1-5\,$d (lower end of group~C in Fig.~2).  
Conventional evolutionary
scenarios suggest that systems where the donor has a
well-developed degenerate core should
inevitably evolve to longer orbital periods.  
However, many of the systems
in group~C of Fig.~2 have not yet developed such cores.  
Moreover, we note
that, of the 20 binary radio pulsars in the Galactic disk population that 
are supposed to fit this evolutionary scenario involving stable mass transfer
from a low-mass giant to the NS, 9 systems have orbital periods shorter 
than $5\,$d, with some $<1\,$d (Rappaport et al.\ 1995).  We suggest that 
detailed binary evolution calculations of these types of systems be 
undertaken.

Our results predict the existence of a large number of binary pulsars
with companion masses $m_2\sim0.05\,M_\odot$ and orbital periods
as short as $\sim15\,$min that may have so far escaped detection
(lower end of group~A in Fig.~2).
Future observations using more sophisticated acceleration-search
techniques or shorter integration times may be able to detect
them (see Camilo et al.\ 2000). They
should approximately follow a period -- companion mass relation given
by $P_b ({\rm d}) \simeq 10^{-5} (m_2/M_\odot)^{-2.5}$.
We also find a large number of post-CE NS-WD binaries with periods
$P_b\sim 1-30$d and WD masses above $0.5\,M_\odot$
(CO WDs; right side of group~B in Fig.~2). No such system
has been definitely observed among the binary radio pulsars in 47~Tuc.
One obvious reason may be that
the NS was not recycled during the short CE phase, altough we must
point out that two systems of this type may have been observed in the 
Galactic disk population (PSR J2145$-$0750 and PSR B0655$+$64;
see, e.g., Phinney \& Kulkarni 1994).

\acknowledgments
We are grateful to F.~Camilo for many useful discussions and for communicating
results in advance of publication. We also thank K.~Joshi and V.~Kalogera
for useful comments. 
This work was supported by NSF Grant AST-9618116 and NASA ATP
Grant NAG5-8460 (to F.A.R.) and NASA ATP Grants NAG5-4057 and NAG5-8368
(to S.R.).
F.A.R.\ was supported in part by an Alfred P.\ Sloan Research Fellowship.
Our computational work is supported by the National Computational Science
Alliance under grant AST980014N.

\end{document}